\documentclass[aip,rsi,reprint]{revtex4-1} 
\usepackage{hyperref}
\usepackage{graphicx,color,amsmath,amssymb,chngpage,mathtools,subfigure}

\definecolor{mygrey}{rgb}{0.5,0.5,0.5}
\definecolor{mygreen}{rgb}{0,0.5,0}
\definecolor{mylightgreen}{rgb}{0,0.75,0}
\definecolor{myblue}{rgb}{0,0,0.75}
\definecolor{mymagenta}{cmyk}{.2,1,0,0.12}
\definecolor{myred}{rgb}{0.95,0,0}

\draft 

\begin{document}


\title{Shot-noise-limited magnetometer with sub-pT sensitivity at room temperature} 




\author{Vito Giovanni Lucivero}
\email[Corresponding author  ]{vito-giovanni.lucivero@icfo.es}
\affiliation{ICFO -- Institut de Ciencies Fotoniques,
Mediterranean Technology Park, 08860 Castelldefels, Barcelona,
Spain}
\author{Pawel Anielski}
\author{Wojciech Gawlik}
\affiliation{Center for Magneto-Optical Research Institute of
Physics, Jagiellonian University Reymonta 4, 30-059 Krakow,
Poland}
\author{Morgan W. Mitchell}
\affiliation{ICFO -- Institut de Ciencies Fotoniques,
Mediterranean Technology Park, 08860 Castelldefels, Barcelona,
Spain} \affiliation{ICREA -- Instituci\'o Catalana de Recerca i
Estudis Avan\c{c}ats, 08015 Barcelona, Spain}


\date{\today}

\begin{abstract}
We report a photon shot-noise-limited (SNL) optical magnetometer
based on amplitude modulated optical rotation using a
room-temperature $^{85}$Rb vapor in a cell with anti-relaxation
coating. The instrument achieves a room-temperature sensitivity of
$70$ fT/$\sqrt{\mathrm{Hz}}$ at \mbox{$7.6$ $\mu$T}. Experimental
scaling of noise with optical power, in agreement with theoretical
predictions, confirms the SNL behaviour from \mbox{$5$ $\mu$T} to
\mbox{$75$ $\mu$T}. The combination of best-in-class sensitivity
and SNL operation makes the system a promising candidate for
application of squeezed light to a state-of-the-art atomic sensor.
\end{abstract}


\maketitle 

\section{Introduction}
\label{sec:Intro} Optical magnetometers
\cite{Budker2002,Budker2007,BudkerBookMagnetometry} are currently
the most sensitive devices for measuring low-frequency magnetic
fields and have  many applications, from medical diagnostics and
biomagnetism \cite{Bison2009,Knappe2010,Johnson2010}, to the
detection of fields in space \cite{Ledley1970,WichtCLEO}, to tests
of fundamental physics
\cite{Budker2006,Romalis2011a,PospelovPRL2013}. Quantum-enhanced
sensitivity of optical magnetometers has been recently
demonstrated using squeezing \cite{ShahPRL2010,WolfgrammPRL2010,
Horrom2012}. Quantum enhancement of a best-in-class magnetometer,
i.e. of an instrument with unsurpassed sensitivity for a given
parameter range, is a natural next step after these
proof-of-principle demonstrations.  This kind of enhancement was
recently shown in gravitational wave detection, when the LIGO H1
detector was enhanced with squeezed light \cite{AasiJ20132}.

In this work we demonstrate a shot-noise-limited magnetometer that
simultaneously is well-adapted for sensitivity enhancement with
squeezed  light, as in \cite{WolfgrammPRL2010, Horrom2012}, and
has detection noise of $70$ fT$/\sqrt{\rm Hz}$ at a field of $7.6$
$\mu$T. For the given field strength and room-temperature atomic
density of $n=1.27\times10^{10}$ atoms/cm$^{3}$ \quad
\cite{Steck}, this is among the best reported magnetometer
sensitivity including those using amplitude
\cite{Higbie2006,Pustelny2008}, frequency
\cite{Acosta2006,Kimball2009} and polarization \cite{Breschi2014}
modulation strategies. With  two orders of magnitudes higher
atomic density, a heated-cell scalar magnetometer (cell
temperature 160$^\circ$ C) showed a noise level below $10$
fT$/\sqrt{\rm Hz}$ in the same field range \cite{Smullin2009}.
Sub-femtotesla spin-exchange-relaxation-free (SERF) magnetometers,
e.g. \cite{Kominis2003,Sheng2013}, are not comparable here because
they operate only at near-zero field.

Our magnetometer is based on the process of nonlinear
magneto-optical rotation (NMOR), also known as nonlinear Faraday
rotation \cite{Gawlik1974,Budker2002}.  In this process, resonant
or near-resonant light produces spin coherence by optical pumping,
and the spin coherence in turn produces Faraday rotation, either
of the optical pumping beam itself \cite{Gawlik2006}, or of a
separate probe beam \cite{Higbie2006}, leading to a detectable
signal indicating the Larmor frequency and thus the magnitude of
the field. Modulation of the pumping, either in frequency (FM
NMOR) \cite{Budker2002a}, amplitude (AMOR) \cite{Gawlik2006} or
circular polarization \cite{Breschi2014} produces a resonant
buildup of atomic polarization with minimal disturbance to the
spin coherence. The modulation strategy significantly increases
the magnetic dynamic range, i.e., the ratio between the largest
detectable signal, which in NMOR can reach the geophysical field
range \cite{Acosta2006,Patton2012}, and the lowest detectable
signal. NMOR can give high sensitivity, due to the long
ground-state coherence times, and hence narrow resonances, that
arise when alkali vapours are confined with a buffer gas
\cite{Brandt1997,Novikova2005} or in anti-relaxation coated cells
\cite{Budker1998,Corsini2013}.

The sensitivity of optical magnetometers is ultimately limited by
two fundamental noise sources: the atomic projection noise and the
optical shot-noise \cite{Budker2002,Budker2007}. When atomic
projection noise is limiting, quantum non-demolition measurement
\cite{KoschorreckPRL2010a, KoschorreckPRL2010b, ShahPRL2010,
SewellNP2013}, atomic entanglement \cite{Wasilewski2010} and spin
squeezing \cite{SewellPRL2012} can improve sensitivity for
measurements within the atomic coherence time \cite{ShahPRL2010}
and for non-exponential relaxation processes
\cite{VasilakisPRL2011}. Similarly, optical squeezing can improve
sensitivity when photonic shot noise is limiting
\cite{WolfgrammPRL2010, Horrom2012}. Prior works on AMOR
\cite{Pustelny2008} and FM NMOR \cite{Kimball2009} have shown
experimental sensitivity about one order of magnitude above (i.e.
worse than) the predicted fundamental sensitivity. Other
magnetometers based on oscillating field-driven Zeeman resonance
\cite{Smullin2009}, M$_{x}$ method \cite{Schultze2010} or
intensity-modulated (IM) pumping \cite{Schultze2012}, have
approached the photon shot-noise level, but still have a
significant technical noise component. In contrast, we report an
AMOR magnetometer in which all other noise sources are
significantly below shot noise from $5$ $\mu$T to $75$ $\mu$T, as
needed for squeezed-light enhancement. We make a detailed and
redundant analysis of the quantum versus classical noise
contributions, including both theoretical calculation of the
expected shot noise level and an independent, fully experimental
analysis based on scaling of measured noise with optical power.
These agree and indicate the potential to improve the sensitivity
of this system by up to 6 dB using polarization squeezing.

The paper is organized as follows: in Section \ref{sec:Setup} we
describe the experimental setup; in Section \ref{sec:Signal} we
explain the modulation strategy  by showing representative AMOR
signals and we define the magnetometer sensitivity. In Section
\ref{sec:results} we describe the optimization of the experimental
parameters to maximize the sensitivity and we present its trend
versus probe light power. In Section \ref{sec:shot} we study the
noise properties of the detection system and we experimentally
demonstrate the shot-noise-limited (SNL) performance of the
optimized magnetometer, by showing agreement with theoretical
predictions.

\begin{figure}[t]
\hspace{-0.4cm}
{\includegraphics[width=\columnwidth]{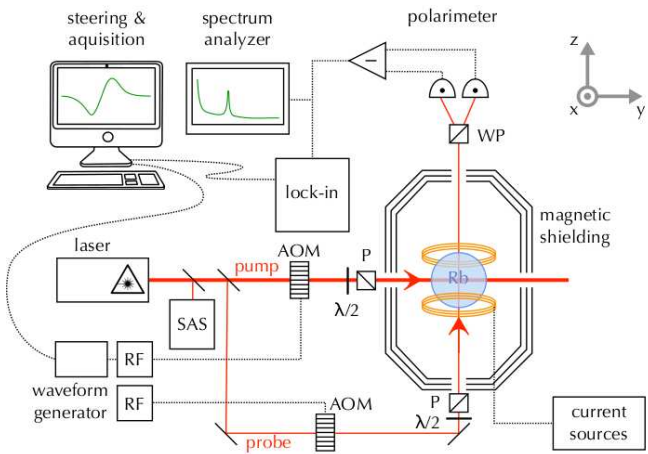}}
\caption{\textbf{Experimental Setup}. SAS --
saturated-absorption-spectroscopy frequency reference, AOM --
acousto-optical modulator with the RF driver, $\lambda/2$ --
half-wave plate, P -- polarizer, WP -- Wollaston prism. The
oscillator that drives the AOM of the pump beam is amplitude
modulated with a sine-wave of frequency $\Omega_m/2\pi$ by the
waveform generator.} \label{setup}
\end{figure}

\section{Experimental Setup}
\label{sec:Setup} The experimental scheme is shown in Fig.
$\ref{setup}$. A sample of isotopically-pure $^{85}$Rb is
contained in a spherical vapor cell of 10 cm diameter, with no
buffer gas. The cell is at room temperature ($\sim25^{\circ}$C)
corresponding to $^{85}$Rb atomic density of $n=1.27\times10^{10}$
atoms/cm$^3$ \cite{Steck}. The inner cell walls are coated with an
antirelaxation (paraffin) layer that prevents atoms from
depolarizing upon collision with the walls and prolongs the
ground-state Zeeman coherence lifetime to $\simeq100$ ms.

The cell is inside a ``box solenoid,'' a cubical box made of
printed-circuit-board  material, with three mutually perpendicular
sets of printed wires, each in a solenoidal pattern. Together with
an accompanying ferrite box, which extends the effective length of
the solenoid based on the method of images for magneto-statics, we
can generate a uniform field along the three directions. In this
experiment we generate a constant magnetic field along the
$z$-axis, which is also the probe beam direction, while the coils
in the perpendicular directions are used to compensate the
residual transverse magnetic field. Residual magnetic field
gradients are compensated by a set of three mutually perpendicular
anti-Helmoltz coils wound around the box. This setup was kept
inside three nested layers of $\mu$-metal shields, giving a whole
magnetic shielding of $\sim 10^6$ efficiency.

The light source for both probing and pumping is an
extended-cavity diode laser whose frequency is stabilized by
saturated absorption spectroscopy at $20$ MHz below the $F=3
\rightarrow F'=2$ transition of the $^{85}$Rb D$_{1}$ line (see
Section \ref{sec:results}). The laser beam is split into pump and
probe beams that pass through acousto-optic modulators
independently driven by two $80$ MHz RF signals so that, before
reaching the atoms, the frequency is additionally red-detuned $80$
MHz away from the $F=3 \rightarrow F'=2$ transition. Additionally,
the intensity of the pump beam is sinusoidally modulated with
frequency $\Omega_{m}/2\pi$. \footnote{The AOM along the probe
beam path makes the setup suitable also for single-beam NMOR but
is not necessary in the strategy followed in this paper, where
just the pump beam needs to be amplitude modulated.} Both pump and
probe have a beam diameter of $1$mm at the center of the vapor
cell.

Both beams are vertically polarized (x-direction in Fig.
\ref{setup}) with high-quality crystal polarizers to ensure pure
linear polarization and the light intensity that interacts with
the atoms can be adjusted with half-wave plates situated in front
of the polarizers. The pump beam passes through the cell in the
$y$ direction, perpendicular to the $z$-axis bias field. When the
pumping modulation frequency coincides with twice the Larmor
precession frequency, a large precessing alignment accumulates in
the $x-y$ plane (see Section \ref{sec:Signal}). The pump power is
set to $60\mu$W. The probe beam propagates through the atomic
vapor cell along the $z$-axis, i.e. parallel to the field, and
experiences Faraday rotation (NMOR) of the polarization plane due
to the precessing alignment. The optimal probe power changes from
$80.5$ $\mu$W to $620$ $\mu$W, depending on the employed
detector's gain (see section \ref{sec:shot}).

Polarization rotation is detected with a balanced polarimeter
consisting of a Wollaston prism set at an angle of $45^{\circ}$
with respect to the vertical and a fiber-coupled variable gain
balanced photo-detector (PDB) (Thorlabs PDB150A DC). The
differential output is analyzed with a radio-frequency (RF)
spectrum analyzer (SA) (RIGOL DSA1030A) or demodulated at
$\Omega_m/2\pi$ with a lock-in amplifier (Stanford Research
Systems model SR844). The in-phase and quadrature output signals
are then stored on a computer for later analysis. As explained in
section \ref{sec:results}, both SA and lock-in signals are used to
determine the magnetometer sensitivity. Throughout this work we
used SA resolution bandwidth RBW$=30$ Hz and video bandwidth
VBW$=30$ Hz.

\newcommand{\Speak}{S_{\rm sig}}
\newcommand{\Sback}{S_{\rm bg}}
\newcommand{\gdet}{g_{\rm det}}

\section{Faraday rotation signal}
\label{sec:Signal} 
\begin{figure}[t]
 \centering
\hspace{-0.2cm}
{\includegraphics[width=\columnwidth]{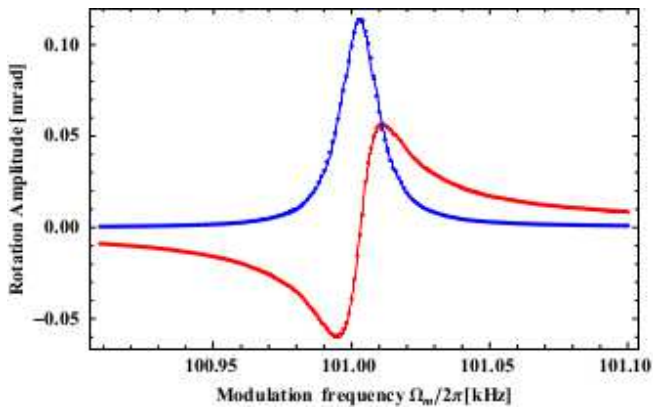}}
\caption{\textbf{AMOR Signals versus Modulation Frequency}.
In-Phase $\phi_\textrm{P}$ (blue) and quadrature $\phi_\textrm{Q}$
(red) output signals of the lock-in amplifier for $B=10.8\mu$T,
$P_{probe}=80\mu$W and $P_{pump}=60\mu$W. The
modulation/demodulation frequency $\Omega_m/2\pi$ is scanned
around the resonance condition $\Omega_m=2\Omega_L$($\Delta=0$).
Experimental data are fitted by dispersive (red) and absorptive
(blue) Lorentzian curves. From the fit we obtain resonance
frequency and FWHM width $\gamma=\Gamma/2\pi$.} \label{fig:Lockin}
\end{figure}

The AMOR signal is generated by means of
amplitude modulated pumping and unmodulated CW probing in a
right-angle geometry. Optical pumping with linearly polarized
light generates spin alignment, i.e. ground state coherences
between Zeeman sub-levels with $\Delta m_F=2$ \quad
\cite{Cohen1974,Cohen1975}. The alignment describes a preferred
axis, but not a preferred direction along this axis. The signal
due to alignment oscillates at twice the Larmor frequency due to
this additional symmetry, i.e., at $2 \Omega_L=
2{g_F\mu_0B}/{\hbar}$ where $g_F$ is the Land\'{e} factor and
$\mu_0$ is the Bohr magneton. Amplitude modulated optical pumping
at $2 \Omega_L$ produces a resonant build-up of spin alignment, as
demonstrated in several earlier works
\cite{Higbie2006,Patton2012}. The alignment behaves as a damped
driven oscillator, and in steady state responds at frequency
$\Omega_m$ with an amplitude and phase relative to the drive that
depend on the detuning $\Omega_m - 2 \Omega_L$ \cite{Wlodar2012}.
The weak probe is sensitive to alignment through linear dichroism,
i.e., linearly polarized light parallel to the alignment
experiences less absorption \cite{BudkerBookMagnetometry}.  When
the alignment is neither parallel to nor perpendicular to the
probe polarization, this dichroism rotates the probe polarization.
This rotation signal also oscillates at $2 \Omega_L$, and we
demodulate it with the lock-in amplifier to extract the in-phase
and quadrature components, shown in Fig. \ref{fig:Lockin} for a
representative magnetic field intensity of $B=10.8$ $\mu$T.
\newcommand{\RBW}{B_{\rm res}}
\renewcommand{\RBW}{{\rm RBW}}
\newcommand{\SNR}{R_{\RBW}}
\renewcommand{\SNR}{{\rm SNR}}
\newcommand{\rms}{{\rm RMS}}
\newcommand{\phimax}{\phi_0}
\newcommand{\phisine}{\phi_s}
\newcommand{\Flux}{\Phi_{\rm ph}}
\newcommand{\myPhi}{\Phi}

The optical rotation angle is an oscillating function at the
modulation frequency $\Omega_m$ with the amplitude dependence well
described by a single-Lorentzian in the small field approximation
\begin{eqnarray}
\phi(t) &=& \phimax {\rm Re}  \left[ \frac{i \Gamma/2}{\Delta + i
\Gamma/2} e^{i \Omega_m t} \right] + \delta \phi(t)  \nonumber \\
&=& \phi_\textrm{P}\cos\left(\Omega_m t\right) +
\phi_\textrm{Q}\sin\left( \Omega_m t\right) + \delta \phi(t)
\label{eq:RotationAngle}
\end{eqnarray}
where $\phimax$ is the maximum rotation angle, which depends on
the optical detuning, cell dimension, and pump power. The detuning
between the modulation frequency and $2\Omega_L$ is $\Delta \equiv
\Omega_m - 2 \Omega_L$ while $\Gamma$ is the FWHM line width due
to relaxation, pumping, and nonlinear Zeeman shifts \footnote{The
single-Lorentzian approximation should fail at large $B$, when the
resonance splits into several lines due to the nonlinear Zeeman
shift. This was not observed at the field strengths used in this
work.  Even at $75 \mu$T, the response was well approximated as a
single Lorentzian. This suggests a strong line-broadening
accompanied the nonlinear Zeeman shift.}. The symbols
$\phi_\textrm{P}$ and $\phi_\textrm{Q}$ are the in-phase and
quadrature components, respectively, directly observable by
demodulation at $\Omega_m$. The photon shot noise contribution,
$\delta \phi(t)$, is a white noise with a power spectral density
$S_\phi(\omega) = 1/(2\Flux)$
\cite{Kimball2009,Pustelny2008,BudkerBookMagnetometry},  where
$\Flux$ is the flux of photons arriving to the detector.

\begin{figure}[t]
 \centering
\hspace{-0.2cm}
{\includegraphics[width=\columnwidth]{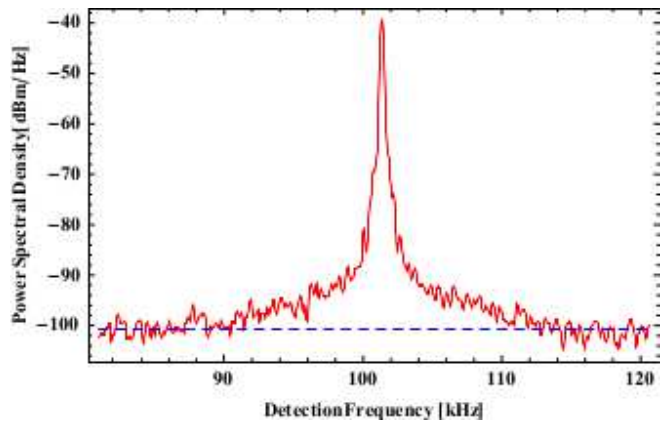}}
\caption{\textbf{AMOR Magnetometer Resonance Spectrum}. Spectrum
of the rotation signal acquired on SA at the resonance condition
$\Omega_m=2\Omega_L$ with RBW$=30$Hz, VBW$=30$Hz and a PDB nominal
gain G$=10^6$V/A. The red curve shows the signal spectrum
$S(\Omega)\equiv \Speak $ with a magnetic field of $B=10.8\mu$T
and $40$kHz span frequency around $\Omega_m$, while the blue
dashed line indicates the background noise level, i.e.
$S(\Omega)\equiv \Sback$ acquired with $B=0$ and averaged over a
$4$kHz range around $\Omega_m$.} \label{fig:peak}
\end{figure}

We note that on resonance, i.e. with $\Delta = 0$, the signal
consists of a cosine wave at frequency $\Omega_m$ with amplitude
$\phimax$, plus a white-noise background due to $\delta \phi(t)$.
In the balanced condition, and with $\phimax \ll \pi$, the
polarimeter signal is $\propto \phi(t)$.  When recorded on a
spectrum analyzer with resolution bandwidth $\RBW$, the signal
shows a peak power spectral density $\Speak = \gdet^2
\phimax^2/(2\RBW)$, where $\gdet$ is the gain relating rotation
angle to RF amplitude at the SA (the factor of one half represents
a mean value of $\langle \cos^2 \rangle = 1/2$). A typical RF
spectrum of the AMOR resonance recorded in our measurements is
shown in Figure \ref{fig:peak}. The signal peak rises above a flat
background $\Sback = \gdet^2 \overline{\delta\phi}^2/2$, where
$\overline{\delta\phi}^2 = S_\phi$ is the spectral noise density
of the phase, so that $\overline{\delta \phi}$ has units
rad/$\sqrt{\rm Hz}$ (the factor of two reflects the fact that only
one quadrature contributes to the noise of the demodulated signal,
while both are recorded by the SA). The signal-to-noise ratio
$\SNR$ is given by $\SNR^2 \equiv \RBW \Speak/\Sback =
\phimax^2/\overline{\delta\phi}^2$, which is independent of
$\gdet$ and $\RBW$ and can be directly measured.

The magnetic sensitivity can be related to $\SNR$ by noting that
the slope of the quadrature component on resonance is
\begin{equation}
\frac{d\phi_\textrm{Q}}{dB} = \frac{g_F \mu_0}{\pi \hbar}
\frac{\phimax}{\gamma}. \label{eq:slope}
\end{equation}
where the width $\gamma\equiv\Gamma/2\pi$ has unit of Hz.
Considering that on resonance
$\Omega_m=2\Omega_L=2{g_F\mu_0B}/{\hbar}$, we find the noise in
magnetic units, i.e., the sensitivity
\begin{eqnarray}
\centering \delta B &=&
\left|\frac{d\phi_\textrm{Q}}{dB}\right|^{-1}
\overline{\delta\phi} = \frac{\pi\hbar}{g_F\mu_0}
\frac{\gamma}{\SNR} , \label{eq:sens}
\end{eqnarray}
with units T/$\sqrt{\rm Hz}$.

As described in the next section, using this method to measure the
sensitivity we find $\delta B$ as low as $70 $
fT/$\sqrt{\mathrm{Hz}}$. For comparison, the atomic projection
noise contribution to the overall measurement is:
\cite{Budker2002,Budker2007}:
\begin{equation}
\centering \delta B_{at}\simeq \frac{\hbar\pi}{g_F\mu_0}
\sqrt{\frac{\gamma}{N_{\rm at}\Delta\tau}} \label{atnoise}
\end{equation}
where $N_{at}$  is the number of atoms involved in the
measurement. With our cell volume of $4\pi R^3/3$, $R\approx 5$
cm, atomic density $n=1.27\times10^{10}$ atoms/cm$^3$, measured
relaxation rate $\gamma\approx10$ Hz and $\Delta\tau=1$s time of
measurement we find $\delta B_{at}\simeq0.134$
fT/$\sqrt{\mathrm{Hz}}$. This value is two orders of magnitude
lower than the observed sensitivity, justifying our earlier step
of ignoring this contribution. If all other noise sources have
lower amplitude than the shot noise, then the magnetometer can be
expected to be photon shot-noise-limited. In Section
\ref{sec:shot} we demonstrate that, in the experimental conditions
that optimize the sensitivity, this is indeed the case.

\section{Optimization of the Magnetometer sensitivity}
\label{sec:results} In this section we examine different setup
parameters in order to find the optimal conditions maximising the
magnetometric sensitivity.

In our configuration, with a pump and probe of the same frequency,
laser tuning affects the pumping efficiency, the rotation signal
corresponding to a given degree of atomic alignment, and the probe
absorption.  In addition, the pump power increases both the
amplitude and the width of the rotation signal. To optimize these
parameters, we first adjust the gradient fields to minimize the
broadening due to magnetic field inhomogeneities
\cite{Pustelny2006}, and then optimize the laser frequency and
pump power to maximize the slope of the AMOR signal. The optimum
conditions, which we use throughout this work, occur at the
detuning of $100$ MHz to the red of the $F=3 \rightarrow F'=2$
transition and $60$ $\mu $W of pump power.

\begin{figure}[t]
\begin{subfigure}
 \centering
\hspace{-0.2cm}
{\includegraphics[width=\columnwidth]{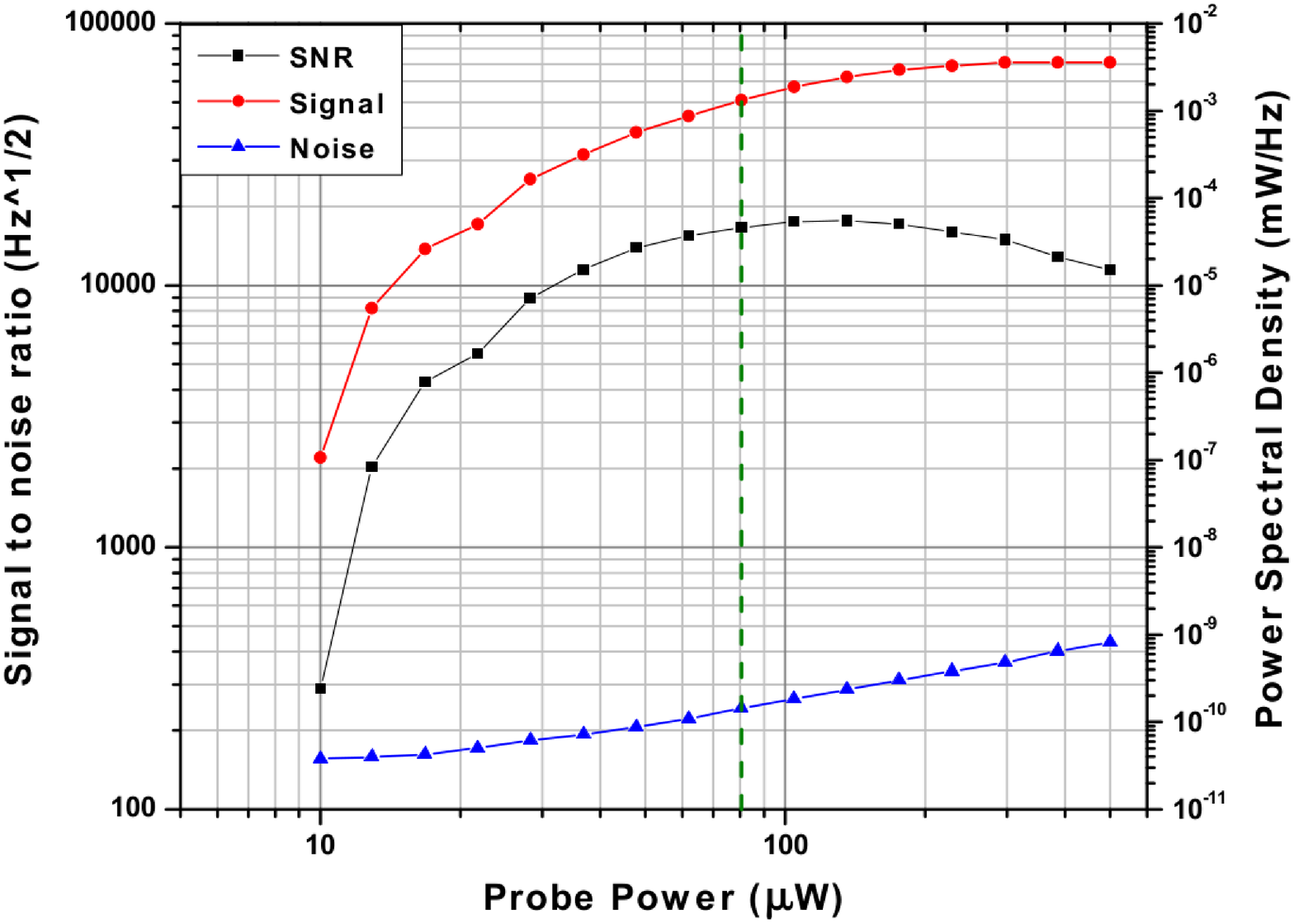}}
 \caption{\textbf{Magnetometer SNR}. Signal-to-Noise ratio versus optical probe power.
The modulation frequency was $71$ kHz ($B=7.6$ $\mu$T). The green
dashed line indicates the probe power value of $80.5$ $\mu$W that
maximizes the sensitivity. This condition does not correspond to
the best SNR because of the trade-off with the width trend (see
Eq. \ref{eq:sens}).} \label{fig:71kHzAmpSNR}
\end{subfigure}
\begin{subfigure}
 \centering
\hspace{-0.2cm}
{\includegraphics[width=\columnwidth]{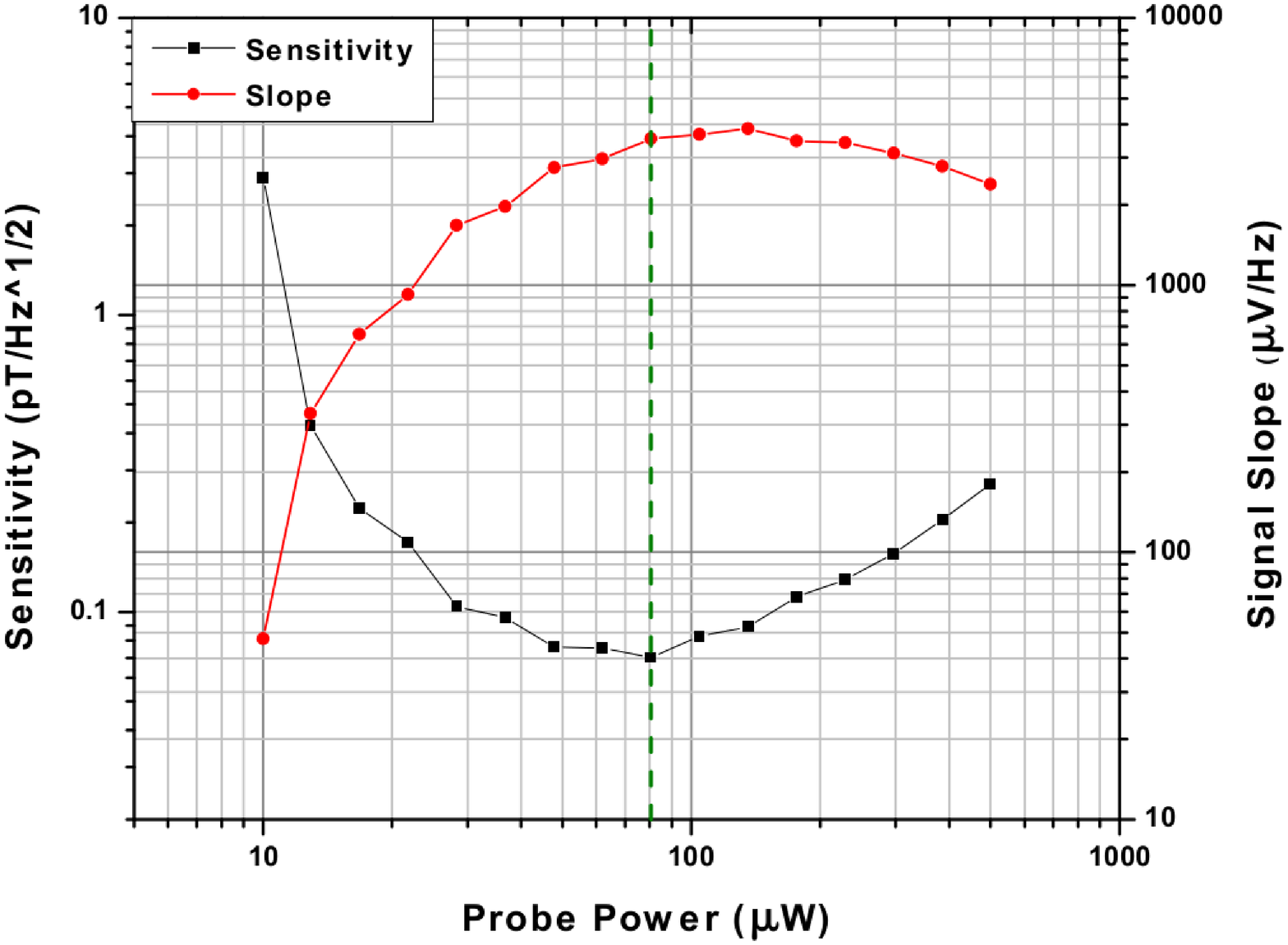}}
 \caption{\textbf{Magnetometer Sensitivity}. Signal slope $\phi_0/\gamma$ and magnetometer sensitivity versus optical probe
 power. The sensitivity is computed as in Eq. \ref{eq:sens} using the width from the demodulated signal, as in Fig.
\ref{fig:Lockin}, and the measured SNR, as in Fig. \ref{fig:peak}.
The green dashed line indicates the probe power that gives the
best sensitivity of $70$ fT$/\sqrt{\mathrm{Hz}}$ for a modulation
frequency of $71$ kHz ($B=7.6$~$\mu$T)}.
\label{fig:71kHzSlopeSensitivity}
\end{subfigure}
\end{figure}

To measure the magnetometric sensitivity for a given probe power
and field strength, we first set the detuning and pump power to
the optimal values discussed above. We then set a constant current
in the solenoidal coil along the $z$-axis, and minimize the width
of the AMOR resonance with the help of the gradient coils.
Demodulation of the signal yields the in-phase and quadrature
components of the resonance versus $\Omega_m$, as depicted in Fig.
\ref{fig:Lockin}. By fitting a Lorentzian to these curves, the
central resonance modulation frequency $\Omega_m=2\Omega_L$
($\Delta=0$) and width $\gamma$ are obtained. Keeping then
$\Omega_m$ fixed and maximizing the in-phase component allows one
to measure the spectrum as in Fig. \ref{fig:peak} and to extract
$\Speak(\Omega_m)=\phimax^2/\RBW$. A second spectrum is taken with
the B-field set near zero.  This moves the resonance peak far away
from $\Omega_m$, so that $S(\Omega_m)$ now gives the background
noise $\delta \phi_{RMS}^2$. In analogy with previous works
\cite{Pustelny2008,Groeger2005} the experimental sensitivity,
defined by equation \ref{eq:sens}, can be calculated in terms of
the width (FWHM) and signal-to-noise ratio. The magnetometric
sensitivity of the instrument was measured in the range from $5$
$\mu$T to $75$ $\mu$T. We employ two detector bandwidths, $300$
kHz and $5$ MHz, corresponding respectively to nominal
transimpedance gains of $10^6$ V/A and $10^5$ V/A.

Typical results, taken at a field of 7.6 $\mu$T (modulation
frequency of $71$ kHz, detector gain setting $10^6$ V/A) are shown
in Figs.  \ref{fig:71kHzAmpSNR} and
\ref{fig:71kHzSlopeSensitivity}. In Fig. \ref{fig:71kHzAmpSNR} we
present signal $\Speak$ and noise $\Sback$ power spectral
densities with the resulting signal-to-noise ratio (SNR) as a
function of the probe power. Signal grows with the probe power
until saturation occurs. In contrast, noise grows monotonically,
so that the SNR has an optimal value before the signal saturates.
Fig.
 \ref{fig:71kHzSlopeSensitivity} depicts the slope $\phimax/\gamma$
and the sensitivity $\delta B$, calculated using equation
\ref{eq:sens}, as a function of probe power, also acquired with
$B= 7.6$ $\mu$T. An optimum sensitivity of 70
fT$/\sqrt\mathrm{Hz}$ is observed at a probe power of 80.5
$\mu$W\quad \footnote{Although this probe power exceeded the pump
power of 60 $\mu$W, the increased resonance broadening is
compensated by higher signal amplitude and results in  a better
net sensitivity. Moreover, for the gain setting of $10^5$ V/A we
find the optimal probe power to be as high as 620$\mu$W.} and
remains within 10\% of this value between 50 $\mu$W and 100
$\mu$W\quad \footnote{While optimized sensitivity value of 70
fT$/\sqrt\mathrm{{Hz}}$ is observed below 10.8 $\mu$T at higher
fields this number rises significantly, roughly as $B^4$, reaching
250 pT$/\sqrt{\mathrm{Hz}}$ at 75 $\mu$T ($\Omega_m=700$ kHz). The
observed reduction of the sensitivity for larger fields is related
to the nonlinear Zeeman effect (NLZ)
\cite{BudkerBookMagnetometry,Acosta2006,Patton2012}. Saturation of
the ferrite shielding cube at high fields and high-order magnetic
field gradients that are not compensated in the current
experimental setup could also contribute to the sensitivity
worsening and need further investigation.}.

\section{Shot-noise-limited performance}
\label{sec:shot} Here we report the results of two noise analyses:
the first characterises the probing and detection system, without
an atomic contribution. This was performed by probing at the
optimal laser detuning but with the pump beam off. The second
analysis characterizes the magnetometer under the experimental
conditions that optimize the sensitivity, as described in section
\ref{sec:results}.

\begin{figure}[t]
\begin{subfigure}
 \centering \hspace{-0.2cm}
{\includegraphics[width=\columnwidth]{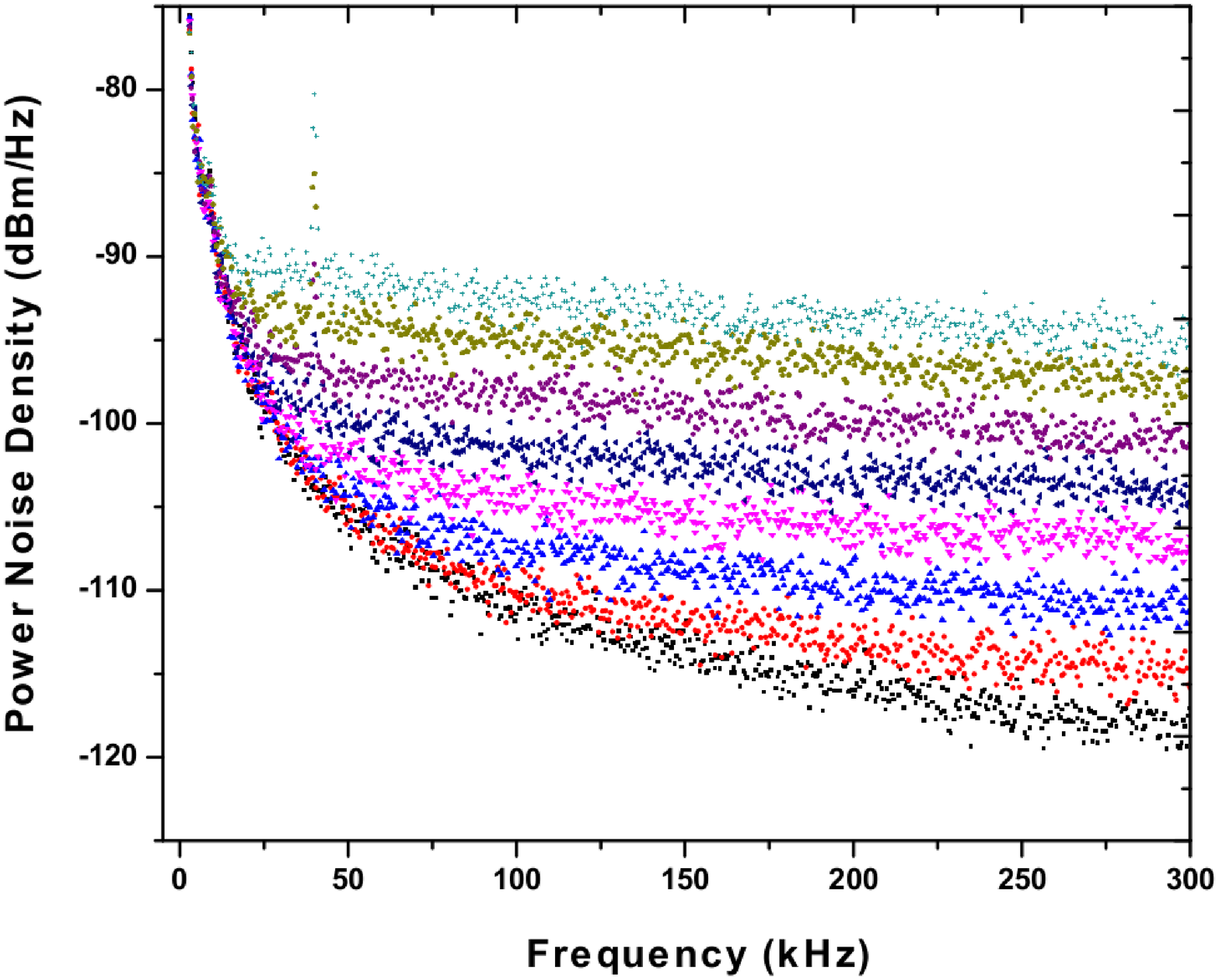}}
\caption{\textbf{Low-Frequency Detection Noise}. Noise spectra of
the PDB differential output acquired with mean optical power of
$P= 0, 10, 20, 50, 100, 200, 400, 700 \mu$W, from bottom to top.
G$=10^6$ V/A and BW$=300$ kHz.} \label{spectra300}
\end{subfigure}
\begin{subfigure}
 \centering
\hspace{-0.2cm}
{\includegraphics[width=\columnwidth]{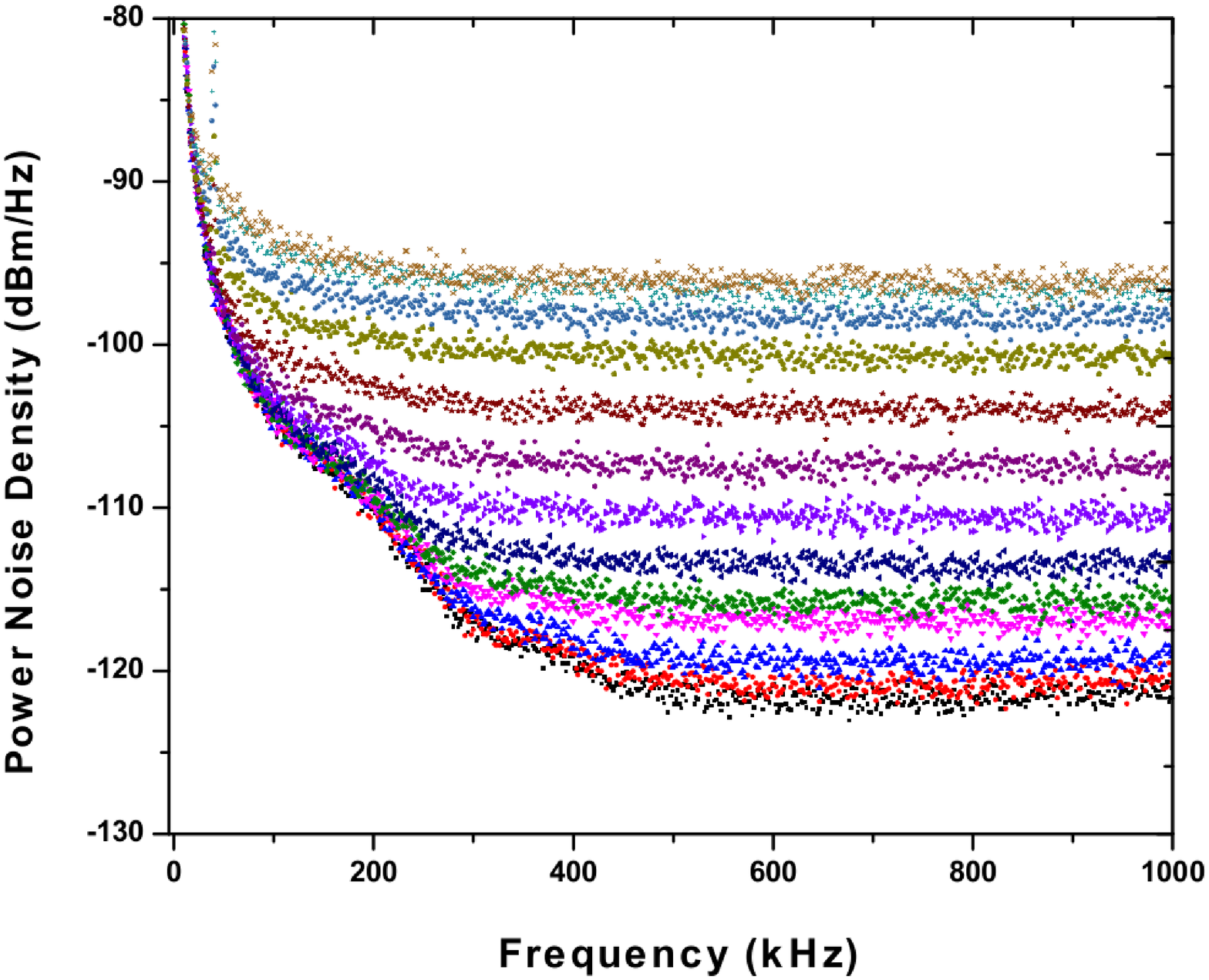}}
\caption{\textbf{High-Frequency Detection Noise}. Noise spectra of
the PDB differential output acquired with mean optical power of
$P= 0, 10, 20, 50, 70, 125, 250, 500, 1000, 2000, 3000, 4500
\mu$W, from bottom to top. G$=10^5$ V/A and BW$=5$
MHz.}\label{spectra5}
\end{subfigure}
\end{figure}

In a linear detection system, the noise power $N$ of the
electronic output will depend on the average light power $P$ as
\begin{equation}
\label{eq:NvsP} N=A P^0+B  P^1+C P^2,
\end{equation}
where $A,B$ and $C$ are constants. The three terms of this
polynomial are the ``electronic noise'' (stemming, e.g. from the
detector electronics), the shot noise, and the ``technical noise''
contributions, respectively \cite{Bachor2004,IcazaAstiz2014}. The
laser source can contribute to technical noise, e.g. through power
fluctuations if the detection is imbalanced or if its optical
elements are unstable. By determining the noise scaling as the
function of light intensity, we can identify the dominating noise
source. When $B P^1>AP^0$ and $B P^1>C P^2$, we say the system is
shot-noise limited, in the sense that the shot noise is the
largest noise contribution.
 These two inequalities define the range of powers
$B/C>P>A/B$ in which the system is SNL.  If $B/C < A/B$, the
system is not SNL for any power. This definition of SNL can be
extended to include more stringent conditions that might arise in
applications.  For any given $k\ge 1$, we can consider powers
satisfying the inequalities $B/(k C) > P> (k A)/B$, i.e. powers
such that the shot-noise contribution is a factor $k$ larger than
both the electronic noise and the technical noise contributions.
For instance, for $k=2$, the photonic noise is $3$ dB higher than
the other two contributions of Eq. (\ref{eq:NvsP}).

For a given field $B$, and thus the Larmor precession or
modulation frequency, the noise of interest is $N = S(\Omega_m)$,
the noise spectral density at the demodulation frequency
$\Omega_m$. Using the SA we collect output noise spectra for
several probe intensities. The data shown in Figs.
\ref{spectra300} and \ref{spectra5} reveal the resulting scaling
of the noise level. The electronic noise floor in the two Figures
differs because of the different employed detector's gain. In the
next step we examine the scaling of the noise level.

For any given detection frequency $\Omega_m$ (that will be the
modulation/demodulation frequency in the magnetometer operation
mode), we can then fit the polynomial of Eq. (\ref{eq:NvsP}), and
find the range of powers and frequencies over which the detection
system is SNL.

\begin{figure}[t]
 \centering
\hspace{-0.2cm} {\includegraphics[width=\columnwidth]{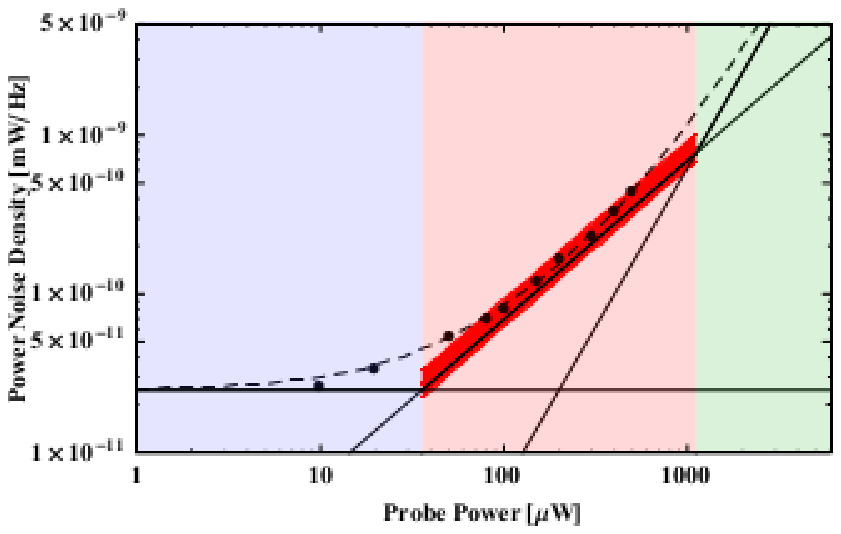}}
\caption{\textbf{SNL Power Range}. At $48.5$ kHz detection
frequency the coefficients of shot-noise (linear scaling) and
technical noise (quadratic scaling) are obtained by fitting data
(black points) with the polynomial function of Eq. \ref{eq:NvsP}
(dashed line), whose intercept is the electronic noise level
(constant), measured at zero power. The red central area
corresponds to the experimental SNL power range. We obtain good
agreement with the theoretical shot-noise level (see text for
calculation) represented by a red line with thickness due to the
10\% uncertainty on the PDB nominal gain $G=10^6$ V/A.}
\label{SNL100}
\end{figure}

In Figure \ref{SNL100} we show an example of this analysis for a
detection frequency of $48.5$ kHz. We can see that scaling of the
noise amplitude is different for different intensity ranges. The
red area represents the SNL range. This is the only power range in
which quantum noise reduction via probe squeezing could
significantly enhance the magnetometer sensitivity. We find good
agreement between the observed shot noise level and the predicted
value  \cite{Bachor2004}  (in W/Hz): $N(\bar{P},\Omega)=
G^2(\Omega) 2\bar{i} e/R $, where $e$ is the electron charge,
$R=50$ Ohm is the SA input impedance and $G(\Omega)$ is detector
gain at frequency $\Omega$. Due to impedance matching at the
detector, the  PDB150A transimpedance gain is only half the
nominal value when used with the SA. The frequency dependence can
be neglected because our signal frequency of $48.5$ kHz is far
below the detector's 300 kHz nominal 3 dB bandwidth. We thus take
$G(\Omega)= (10^6 \pm 10^5)/2$ V/A, which represents the
manufacturer's specification. The photocurrent is
$\bar{i}={\bar{P}e}/({h\nu\eta_{det}})$ where $\bar{P}$ is the
averaged optical power and $\eta_{det} = 0.88$ is the detector
quantum efficiency.

\begin{figure}[t]
\begin{subfigure}
 \centering \hspace{-0.2cm}
{\includegraphics[width=\columnwidth]{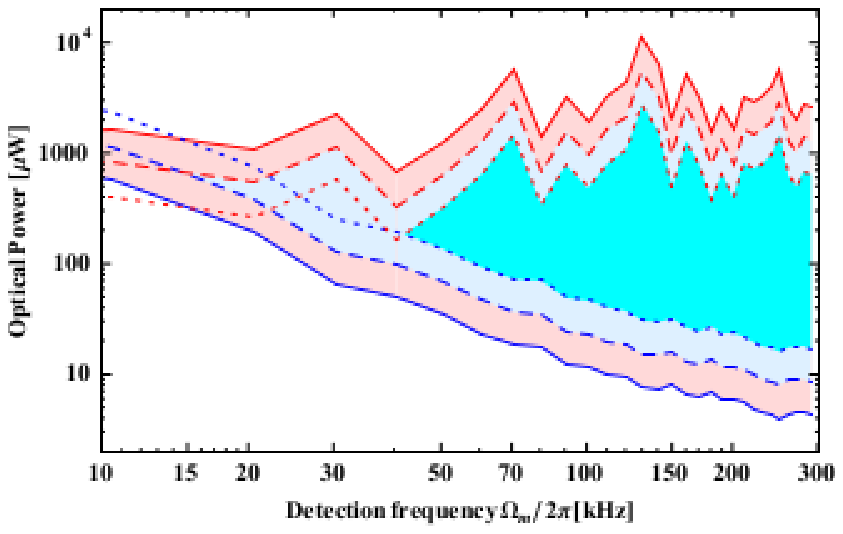}}
\caption{\textbf{SNL Power Range for low frequencies}. Blue and
red curves show $(k A)/B$ and $B/(k C)$, the lower and upper
limits, respectively, of the SNL range with $k=1$ (red region),
$k=2$ (blue region) and $k=4$ (cyan region). PDB gain $G=10^6$
V/A. $A/B$ and $B/C$ were found by fitting the spectra of Fig.
\ref{spectra300} as illustrated in Fig. \ref{SNL100}.  To reduce
scatter, spectra were first averaged in $10$ kHz bins. See text
for details.} \label{SNL300kHz}
\end{subfigure}
\begin{subfigure}
 \centering
\hspace{-0.2cm}
{\includegraphics[width=\columnwidth]{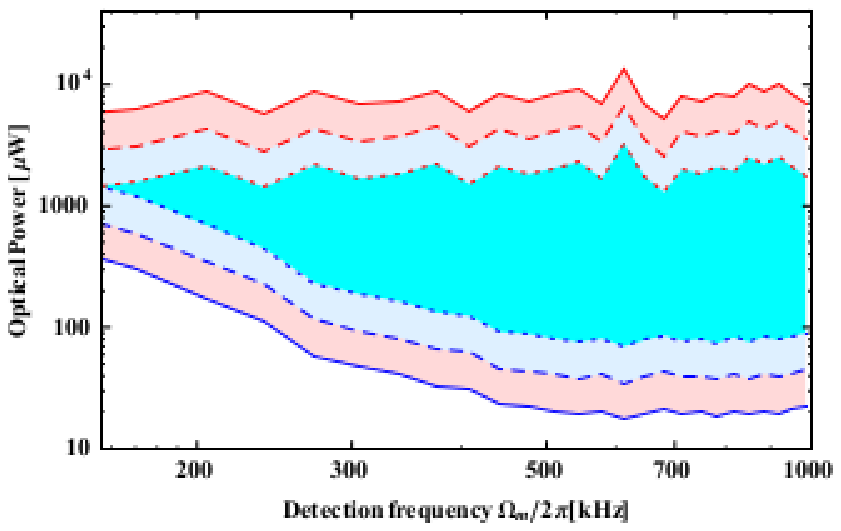}}
\caption{\textbf{SNL Power Range for high frequencies}. Blue and
red curves show $(k A)/B$ and $B/(k C)$, the lower and upper
limits, respectively, of the SNL range with $k=1$ (red region),
$k=2$ (blue region) and $k=4$ (cyan region). PDB gain $G=10^5$
V/A.  $A/B$ and $B/C$ were found by fitting the spectra of Fig.
\ref{spectra5} as illustrated in Fig. \ref{SNL100}.  To reduce
scatter, spectra were first averaged in $34$ kHz bins. See text
for details.} \label{SNL5MHz}
\end{subfigure}
\end{figure}

After performing the same analysis over all detection frequencies,
we report in Figs. \ref{SNL300kHz} and \ref{SNL5MHz} lower ($A/B$)
and upper ($B/C$) power limits of the SNL range (red area) versus
frequency for two detector settings ($10^6$ V/A with 300 kHz BW
and $10^5$ V/A with 5 MHz BW). According to our previous
definition we also show the signficant SNL regions with $k=2$
(light blue area) and $k=4$ (cyan area) that correspond to power
regions where the photonic shot-noise is respectively $3$ dB and
$6$ dB above the other noise contributions. Below $20$ kHz the
detection system is limited by electronic noise i.e. not
significantly SNL within the investigated range of light power. It
is properly reproduced in Fig. \ref{SNL300kHz}, although the data
coming from the fit procedure suffer from considerable scattering.
The dip in the red curves ($B/kC$) at $40$ kHz is due to technical
noise excess at this frequency (see Fig. \ref{spectra300}).

Being interested in the SNL range, we have constrained our AMOR
measurements, reported in section \ref{sec:results}, to
modulation/detection frequencies higher than $50$kHz and thus to
magnetic field intensities above $5$ $\mu$T. Above modulation
frequency of $200$ kHz higher detector BW of 5 MHz needs to be
used. Because of the lower gain ($10^5$ V/A), starting from
frequency of $200$ kHz the system becomes SNL above $200$ $\mu$W
as shown in Fig. \ref{SNL5MHz}.

Having determined the SNL range of the detector, we now proceed to
characterize the magnetometer noise  over this range. We set
conditions for an optimized B-field measurement, as described in
section \ref{sec:results},  and switch off the $B$-field but leave
on all other components, including the modulated pumping (in
contrast to the measurements described above).  We then acquire
the noise power spectrum $\Sback(\Omega_m)$ as a function of probe
power. We report two representative results that correspond to
detector setting with high gain and low gain respectively.
Although in our experiment we did not observe any significant
difference between the detector and the magnetometric noises,
these two features may differ in other experimental conditions
where environmental or technical noise sources dominate over the
fundamental shot-noise contribution, as reported in previous works
\cite{Pustelny2008,Breschi2014}.

\begin{figure}[t]
 \centering
\hspace{-0.2cm}
{\includegraphics[width=\columnwidth]{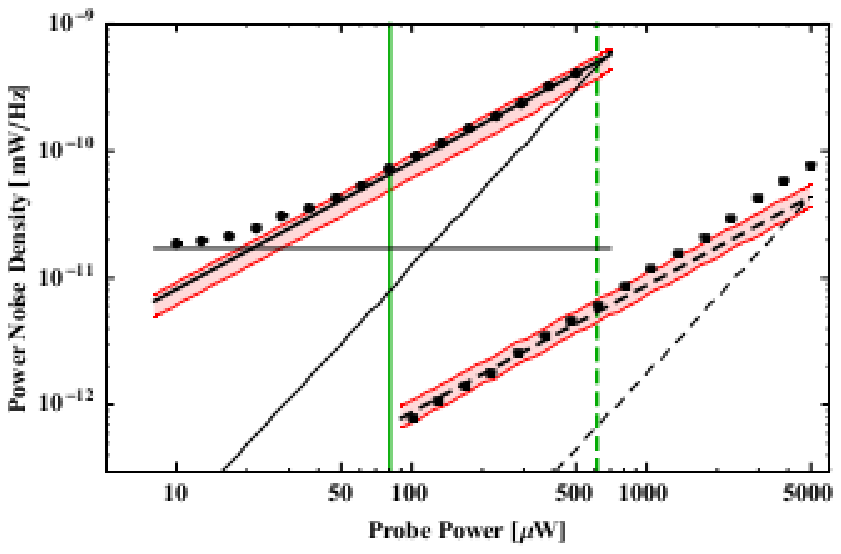}}
 \caption{\textbf{SNL Magnetometer Performance}.
Background noise level (acquired with $B=0$ and averaged over a
$4$ kHz range around the resonance frequency) versus optical probe
power at $71$ kHz (black circles) and $700$ kHz (black squares).
These are simultaneously AM frequencies of the optical pumping
(kept on in the noise measurement) and SA detection frequencies.
Electronic (constant), shot-noise (linear) and technical noise
(quadratic) contributions are shown with solid and dashed black
lines at $71$ kHz and $700$ kHz respectively. For $700$ kHz, the
electronic noise level is below the shown scale at $5.6 \times
10^{-14}$ mW/Hz. Red lines represent the theoretical shot-noise
levels, calculated with the PDB nominal gain values of $G=10^6\pm
10^5$ V/A and $G=10^5\pm 10^4$ V/A respectively. The probe powers
that maximize the magnetometer sensitivity (vertical green lines)
fall within a significant SNL power range. See text for details.}
\label{fig:71SNL}
\end{figure}

In Fig. \ref{fig:71SNL} we show the magnetometer noise power at
$71$ kHz ($B=7.6$ $\mu$T) and $700$ kHz ($B=75$ $\mu$T) as a
function of probe power. Fitting both data with $N(P)$ of Eq.
(\ref{eq:NvsP}) and knowing the electronic noise coefficient $A$,
we find the coefficients $B$ and $C$ and we can define the
shot-noise-limited power range. The difference in power range and
reference level between the two representative frequencies is due
to the different employed BPD gain. The trend of the noise power
is linear, i.e. shot-noise-limited, in the power range of $30$
$\mu$W-$500$ $\mu$W and $100$ $\mu$W$-1$ mW for $71$ kHz and $700$
kHz respectively. Within this range the observed noise levels
agree with theoretical shot-noise levels, calculated in the same
way as for Fig. \ref{SNL100}, by taking into account the 10\%
uncertainty on the detector nominal gain.

Most importantly, the probe power intervals in which the
magnetometer sensitivity is not worse than $10\%$ of the maximum
(reached at $80.5$ $\mu$W and $620$ $\mu$W respectively) are well
inside a significant photon SNL region with $k=4$, in which the
photonic shot-noise is more than $6$ dB above the electronic and
technical noise levels. Within this optimal power range,
significant sensitivity enhancement can be achieved by using
optical polarization squeezing of the probe beam
\cite{Wolfgramm2010}. Indeed, the results of Fig. \ref{fig:71SNL}
show that the fundamental light shot-noise contribution dominates
the magnetometer noise budget i.e. technical noise and atomic
projection noise (Eq. \ref{atnoise}) are negligible when the
magnetometer sensitivity is optimized at room temperature. Similar
SNL performance was observed between $5 \mu$T and $75 \mu$T, over
all the investigated magnetic dynamic range.

\section{Conclusions}
\label{sec:Conclusions} We have demonstrated a sensitive
pump-probe optical magnetometer that is shot-noise limited over
the field range $5$ $\mu$T to $75$ $\mu$T. We optimized the system
for pump/probe detuning, pump and probe beam powers, and found
sensitivity of 70 fT/$\sqrt{\rm Hz}$ at a field of $7.6$ $\mu$T.
The shot-noise-limited performance of the system has been
confirmed by the scaling of the magnetometer noise as a function
of probe input power and by agreement with the theoretical
shot-noise level. This is the first experimental demonstration of
a photon shot-noise-limited AMOR magnetometer. Moreover, it has
the highest reported sensitivity for a room-temperature optical
magnetometer in a range around $10$ $\mu$T. Based on these
observations, the described magnetometer is a good candidate for
squeezed-light enhancement of sub-pT sensitivity over a broad
dynamic range.

It is worth noting that AMOR and other modulated magnetometry
strategies at these field strengths are well-matched to
atom-resonant sources of squeezed light, because the signal is
recovered at a multiple of the Larmor frequency, i.e. at a radio
frequency.  Although optical squeezing can be generated at low
frequencies \cite{VahlbruchPRL2006}, in practice most squeezing
experiments, and to date all atom-resonant squeezed light sources
\cite{TanimuraOL2006, HetetJPB2007, PredojevicPRA2008,
AppelPRL2008, BurksOE2009}, have shown squeezing at radio
frequencies.

A number of improvements suggest themselves.  The lower limit of
$5 \mu$T is set by the low-frequency electronic noise of the
balanced detector.  Electronics designed for lower frequency
ranges \cite{VahlbruchPRL2006} could make the system
shot-noise-limited also for weaker fields.   Recently-developed
anti-relaxation coatings \cite{CorsiniPRA2013} could extend the
ground-state coherence.  Techniques to evade broadening due to the
nonlinear Zeeman effect could improve the sensitivity at high
fields \cite{Acosta2008a,Jensen2009,Chalupczak2010}.

\section*{Acknowledgements}
We thank Adam Wojciechowski, R. Jim\'{e}nez Mart\'{i}nez, F. M.
Ciurana and Y.A. de Icaza Astiz for helpful discussions. This work
was supported by the Spanish MINECO project MAGO (Ref.
FIS2011-23520), European Research Council project AQUMET, Polish
National Science Center (2012/07/B/ST2/00251), and by the
Foundation for Polish Science (TEAM).
\bibliography{AMORSNL}

\end{document}